\documentclass[aps,floatfix,eqsecnum,showpacs,preprint,tightenlines]{revtex4-1}
\usepackage{epsfig}

\usepackage{amsmath}

\begin{document}


\title{Testing semi-local chiral two-nucleon interaction in selected electroweak processes}

\author{R. Skibi{\'n}ski}
\affiliation{M. Smoluchowski Institute of Physics, Jagiellonian University, PL-30059 Krak\'ow, Poland}
\author{J. Golak}
\affiliation{M. Smoluchowski Institute of Physics, Jagiellonian University, PL-30059 Krak\'ow, Poland}
\author{K. Topolnicki}
\affiliation{M. Smoluchowski Institute of Physics, Jagiellonian University, PL-30059 Krak\'ow, Poland}
\author{H. Wita{\l}a}
\affiliation{M. Smoluchowski Institute of Physics, Jagiellonian University, PL-30059 Krak\'ow, Poland}
\author{E. Epelbaum}
\affiliation{Institut f\"ur Theoretische Physik II, Ruhr-Universit\"at Bochum, D-44780 Bochum, Germany}
\author{H. Kamada}
\affiliation{Department of Physics, Faculty of Engineering, Kyushu Institute of Technology, Kitakyushu 804-8550, Japan}
\author{H.Krebs} 
\affiliation{Institut f\"ur Theoretische Physik II, Ruhr-Universit\"at Bochum, D-44780 Bochum, Germany}
\author{Ulf-G.~Mei{\ss}ner}
\affiliation{Helmholtz-Institut~f\"{u}r~Strahlen-~und~Kernphysik~and~Bethe~Center~for~Theoretical~Physics,
~Universit\"{a}t~Bonn,~D-53115~Bonn,~Germany}
\affiliation{Institut f\"ur Kernphysik, Institute for Advanced Simulation
and J\"ulich Center for Hadron Physics, Forschungszentrum J\"ulich,
D-52425 J\"ulich, Germany}
\affiliation{JARA~-~High~Performance~Computing,~Forschungszentrum~J\"{u}lich,~D-52425~J\"{u}lich,~Germany}
\author{A. Nogga}
\affiliation{Institut f\"ur Kernphysik, Institute for Advanced Simulation
and J\"ulich Center for Hadron Physics, Forschungszentrum J\"ulich,
D-52425 J\"ulich, Germany}
\affiliation{JARA~-~High~Performance~Computing,~Forschungszentrum~J\"{u}lich,~D-52425~J\"{u}lich,~Germany}

\date{\today}

\begin{abstract}
The recently developed semi-local improved chiral nucleon-nucleon interaction is used for the first time to study several 
electromagnetic and weak processes at energies below the pion production threshold. 
Cross sections and selected polarization 
observables for deuteron photodisintegration, 
nucleon-deuteron radiative capture, three-body $^3$He photodisintegration 
as well as capture rates for decays of the muonic $^2$H and $^3$He atoms are calculated.
The Lippmann-Schwinger and Faddeev equations in momentum space 
are solved to obtain nuclear states.
The electromagnetic current operator is taken as a single nucleon current supplemented 
by many-body contributions induced via the Siegert theorem.
For muon capture processes the nonrelativistic weak current together 
with the dominant relativistic corrections is used. 
Our results compare well with experimental data, demonstrating the same quality as 
is observed for the semi-phenomenological AV18 potential.
Compared to the older version of the chiral potential with a nonlocal regularization, 
a much smaller cut-off dependence is
found for the state-of-art chiral local interaction employed in this paper.
Finally, estimates of errors due to the truncation of the chiral expansion are given.

\end{abstract}


\maketitle


\section{Introduction}
Studies of electromagnetic and weak reactions are an important part of nuclear physics.
They deliver information on electromagnetic properties of nuclei, transitions between nuclear states
and details of electromagnetic and weak currents inside nuclei 
~\cite{Schiavilla,Bacca,Kammel,Leidemann}.   
In the broader sense the precise description of electromagnetic and weak reactions is a challenging test 
for models of the nuclear Hamiltonian and current operators as well as for the used theoretical schemes and numerical methods.
Chiral Effective Field Theory ($\chi$EFT) is currently the most important theoretical 
approach, which can and should be tested in investigations of electromagnetic and weak processes.
Among many attempts to do that in the few-nucleon sector we mention~\cite{Bacca, Marcucci1, Marcucci2} and references therein.
However, these works were usually restricted to small energies or the lowest orders of the chiral expansion or 
combine the phenomenological potentials and chiral currents within the so-called "hybrid" $\chi$EFT approach
or cover only selected reaction channels.

The continuous progress in the field of $\chi$EFT has resulted in the development of sophisticated  
nucleon-nucleon (NN) 
and many-nucleon interactions~\cite{Machleidt_review,Epelbaum_review}. These forces have been
used to describe reactions in three-nucleon (3N) systems~\cite{Epelbaum,EpelbaumEPJ2002,EpelbaumEPJ2003,GlockleEPJ2004,Kievsky,Viviani}, 
the structure of 
light nuclei~\cite{struktura_chiral} and nuclear matter~\cite{Hebeler}. 
Results obtained for NN scattering at energies up to 300 MeV
and nucleon-deuteron elastic scattering proved the usefulness and high quality of chiral potentials.
In these first studies the regularization of NN and 3N potentials in momentum space 
with the nonlocal regulator was used~\cite{nucleon-nucleon}. 

Recently it was shown~\cite{lenpic1,lenpic_Golak} that such a regularization scheme introduces 
artifacts and affects the correct physical behavior of the potential at long distances.
In addition, the spectral function regularization had to be introduced in order to cut off the unwanted short-range part of
the two-pion exchange potential. This in turn causes too strong dependence of predictions on values
of the cut-off parameters~\cite{Skibinski3H,lenpic1}. 
The Bochum-Bonn group proposed recently in Refs.~\cite{imp1,imp2} an improved
version of the chiral potential up to fifth order of the chiral expansion (N$^4$LO).
During its construction, regularization is performed in coordinate space 
and only afterwards such a regularized force is transformed to momentum space.
In particular, the one-pion and the static two-pion exchange potentials are regularized
in coordinate space by multiplying them by the function $ f(\frac{r}{R})=[1-\exp(-(\frac{r}{R})^2 ]^6 $,
with $R$ being the cut-off (regularization) parameter.  
This procedure maintains the long-range part of the interaction and  
leads to smaller undesirable regularization effects.
Indeed, as shown in~\cite{imp1, imp2} the NN phase shifts as well as the deuteron properties
are much less sensitive to the values of the
regularization parameter $R$ than the ones obtained within the older version~\cite{nucleon-nucleon}.
The same is also true for elastic nucleon-deuteron (Nd) scattering observables~\cite{lenpic3}.
The three-nucleon force with consistent local regularization is under 
development, so in this paper we only use two-body interactions and, in addition, 
we neglect the Coulomb force.  

One of important and unique features of the $\chi$EFT approach
is a possibility of a consistent derivation of nuclear forces and electroweak current operators, see
e.g. Refs.~\cite{Park,Pastore,Baroni,Kolling1, Kolling2} for works in these direction.
In particular, in Ref.~\cite{Kolling1} 
the long-range part of the leading two-pion exchange contributions
was derived. 
Together with the well known single nucleon current (SNC) and consistent old, nonlocal 
version of chiral interaction they were
used to study the deuteron and the $^3$He photodisintegration in Refs.~\cite{Rozpedzik,Skibinski_APP2}.
While the general description of observables was reasonable, a strong cut-off dependence
of predictions was also observed. Similar results were obtained for
radiative nucleon-deuteron capture and $^3$He photodisintegration 
where instead of explicit many-body currents the Siegert theorem was used~\cite{Skibinski_APP1}.
The strong variation of predictions due to different values of the cut-off parameters practically
precluded us from drawing detailed physical conclusions.
Thus it is very interesting to see whether the cut-off dependence also becomes smaller for the electromagnetic processes, 
when the newly developed improved chiral interactions are considered.
In the present work we use the Siegert approach and postpone studies based on explicit single nucleon and many-body 
electromagnetic currents until a more complete picture of the electroweak current operator,
consistent with the NN interaction at each order of the chiral expansion, is known.

The question of the cut-off dependence can be also addressed in weak reactions.
Thus we use the improved NN forces~\cite{imp1, imp2} to calculate the capture rates in muon capture reactions 
on the deuteron and $^3$He.  
In this paper we employ 
a nonrelativistic 
single nucleon weak current operator supplemented with leading relativistic corrections 
(RC)~\cite{Marcucci1,Shen2012,Golak_weak}. Since we are mainly interested in the cut-off dependence of 
capture rates, the use of such an incomplete model of the current operator is justified in the present investigation.

The paper is organized as follows. In the next section we give a short overview of our 
formalism for electromagnetic and weak processes. Section~\ref{chap3} contains selected results for 
the deuteron photodisintegration process, $\gamma + {\rm d} \rightarrow {\rm p} + {\rm n}$, while in Sections~\ref{chap4} and~\ref{chap5}
we discuss 3N electromagnetic processes: nucleon-deuteron radiative capture, ${\rm n(p)} + {\rm d} \rightarrow \gamma + ^3{\rm H(}^3{\rm He)}$,
and the total $^3$He photodisintegration $\gamma + ^3{\rm He} \rightarrow {\rm p} + {\rm p} + {\rm n}$, respectively.
The results for weak muon capture processes, $\mu^-+^2{\rm H} \rightarrow {\rm n} + {\rm n} + \nu_{\mu}$,  
$\mu^-+^3{\rm He} \rightarrow ^3{\rm H} + \nu_{\mu}$, 
$\mu^-+^3{\rm He} \rightarrow {\rm n} + {\rm d} + \nu_{\mu}$ and 
$\mu^-+^3{\rm He} \rightarrow {\rm n} + {\rm n} + {\rm p} + \nu_{\mu}$ are presented in Section~\ref{chap6}. 
We summarize in Section~\ref{chap7}.

\section{Formalism}
The theoretical approach used in the present study is described in detail 
in~\cite{Golak,skib2b,raport2005,metody3,Rozpedzik}, so here we only remind the reader of
the key steps. We work in momentum space and employ a formalism based on the 3N Faddeev equations.
The nuclear matrix elements for electromagnetic or weak disintegration processes are 
the central quantities from which we are able to calculate observables~\cite{raport2005, Golak_weak}.

In the case of the deuteron photodisintegration, the nuclear matrix element $N^{\mu}_{deu}$
is defined as 
\begin{equation}
\label{eq3}
 N^{\mu}_{deu} \equiv \langle \Psi^{2N}_{\rm scatt}|j^{\mu}_{2N}|\Psi^{2N}_{\rm bound}\rangle \,,
\end{equation}
where $|\Psi^{2N}_{\rm scatt}\rangle$ and $|\Psi^{2N}_{\rm bound}\rangle$ are the final proton-neutron 
scattering state and the initial deuteron bound state, respectively.
The deuteron state is a solution of the Schr\"odinger equation with the Hamiltonian comprising the
NN potential $V$. The same interaction, together with the free two-nucleon (2N) propagator $G_0$, 
enters the Lippmann-Schwinger equation for the $t$ operator,
$t = V + tG_0 V,$ which we use to obtain the final scattering state.
Then the $N^{\mu}_{deu}$ is given as
\begin{equation}
\label{eq5}
 N^{\mu}_{deu} = \langle \vec{p_{0}} \mid \left( 1 + t G_0 \right) \, j^{\mu}_{2N}
\mid \Psi^{2N}_{\rm bound}\rangle,
\end{equation}
where $|\vec{p}_{0}\rangle$ is the eigenstate of the relative proton-neutron momentum.
The form of the electromagnetic current operator $j^{\mu}_{2N}
$ is discussed below.

Radiative nucleon-deuteron capture is related via the time reversal symmetry to the two-body $^3$He or $^3$H 
photodisintegration reactions. We exploit this relation and calculate the nuclear matrix element
for the radiative Nd capture $N^{\mu}_{\rm{rad Nd}}$ from the matrix element
$N^{\mu}_{Nd} \equiv \langle \Psi^{Nd}_{\rm scatt} \mid ( 1 + P ) j^{\mu}_{3N}
\mid \Psi^{3N}_{\rm bound} \rangle$ for the photodisintegration reaction, 
leading to the final nucleon-deuteron
scattering state $\mid \Psi^{Nd}_{\rm scatt}  \rangle$~\cite{Golak, raport2005}.
The matrix element $N^{\mu}_{Nd}$ can be expressed as
\begin{equation}
N^{\mu}_{Nd} = \langle \phi_{{\rm Nd}} \vert (1+P) j^{\mu}_{3N} \vert \Psi^{3N}_{\rm bound} \rangle + 
\langle \phi_{{\rm Nd}} \vert P \vert U^{\mu} \rangle  \,,
\end{equation}
where $ \vert \Psi^{3N}_{\rm bound} \rangle$ 
is the 3N bound state while
$\vert \phi_{{\rm Nd}} \rangle$ is the product of the internal deuteron state 
and the state describing the free relative motion of the third nucleon 
with respect to the deuteron.
$P = P_{12} P_{23} + P_{13} P_{23} $ is a permutation operator
with $P_{ij}$ being the operator exchanging nucleons $i$ and $j$.

The auxiliary state $\mid U^\mu \, \rangle $ fulfills the Faddeev-like equation~\cite{raport2005}
\begin{equation}
\label{U3n}
\mid U^\mu \,  \rangle =
t G_0 \, ( 1 + P ) j^{\mu}_{3N}
\mid \Psi^{3N}_{\rm bound} \, \rangle
+ t G_0 P \mid U^\mu \, \rangle  \,,
\end{equation}
with the free 3N propagator $G_0$.

The nuclear matrix element for $^3$He photodisintegration leading to three free 
nucleons in the final state, 
$N^{\mu}_{3N} \equiv \langle \Psi^{3N}_{scatt} \mid j^{\mu}_{3N} \mid \Psi^{3N}_{\rm bound} \rangle \; $,
is also given by the auxiliary state $\mid U^\mu \, \rangle $:
\begin{eqnarray}
N^{\mu}_{3N} &=&
\langle \Phi_{3N} \mid ( 1 + P ) j^{\mu}_{3N}
\mid \Psi^{3N}_{\rm bound} \rangle
+ \langle \Phi_{3N} \mid ( 1 + P ) \mid U^{\mu} \rangle  \, ,
\label{eqU}
\end{eqnarray}
where $\mid \Phi_{3N} \rangle$ is an antisymmetrized state 
describing the free motion of the three outgoing nucleons.

The nuclear electromagnetic current operators, $j^{\mu}_{2N}$ and $j^{\mu}_{3N}$, 
employed for the deuteron and $^3$He ($^3$H) photodisintegration
processes, are constructed in the same way. Unfortunately the 2N currents fully 
consistent with the locally regularized
nuclear potentials are not yet available. Therefore, we approximate the electromagnetic current by
a sum of contributions from the individual nucleons and supplement these results by the many-body parts 
incorporated via the Siegert theorem~\cite{Golak,raport2005}. To this end we perform 
a multipole decomposition of the corresponding single nucleon current matrix elements
and use standard identities \cite{raport2005} to express a part of the electric multipoles by the Coulomb  
multipoles, generated again by the single nucleon charge density operator. 
This step is justified by the fact that at low energies many-nucleon contributions to the nuclear charge density are small.   
The remaining part of the electric multipoles and all the magnetic multipoles are taken solely from the single nucleon current operators.
The Siegert theorem, together with the so-called Siegert hypothesis, is also widely used in 
many calculations performed  in coordinate space, see e.g.~\cite{Schiavilla,Arenhoevel1,Arenhoevel2,Arenhoevel3}.
This form of the nuclear current operator has the (purely technical) advantage that it does not depend on the nucleon-nucleon potential
employed in our calculations. We denote this model of the current operator as SNC+Siegert.

The weak muon capture processes on atomic nuclei are described by means of similar nuclear matrix elements
as in Eqs.~(\ref{eq3})-(\ref{eqU}),
which are combined with the well-known leptonic part~\cite{Walecka}, to build the full transition amplitudes.
Depending on the studied process the
deuteron or $^3$He wave function represents the initial nuclear bound state. 
For the $\mu^-+^2~{\rm{H}} \rightarrow {\rm n} + {\rm n} + \nu_{\mu}$ reaction,
the final two-neutron scattering state is calculated using the $t$-operator.
In the case of break-up channels in muon capture on $^3$He 
the corresponding two- and three-body scattering states are required. We calculate them analogously to the
states for photodisintegration, again using Eq.(\ref{U3n}), with the electromagnetic current
$j^{\mu}_{3N}$ replaced by the weak current $j^{\mu}_{w}$~\cite{Marcucci1,Shen2012,Golak_weak}. Since we solve Eq.(\ref{U3n}) at a given 
energy of the 3N system, numerous solutions of Faddeev-like equation (\ref{U3n}) 
have to be obtained in order to cover the whole range of the final muon neutrino energies.
This significantly increases the numerical complexity
of such calculations. Of course, in the case of $\mu^-+^3{\rm He} \rightarrow ^3{\rm H} + \nu_{\mu}$ channel 
we use only pre-calculated $^3$He and $^3$H states and no Faddeev equation for the bound state 
has to be solved repeatedly.
For the weak current $j^{\mu}_{w}$ we employ a non-relativistic single nucleon current operator 
supplemented by the dominant $(p/M_{nucl})^2$ relativistic corrections,
where $M_{nucl}$ is the nucleon mass.
A detailed discussion of the weak current, formulas connecting the nuclear matrix elements with
the capture rates and various aspects of the reaction kinematics are given in~\cite{Golak_weak}.

Our calculations are performed in momentum space and we use the formalism of partial waves.
In the calculations we employ all partial waves in the two-body systems up 
to the two-body total angular momentum j$\leq$3 and in the three-body 
states up to the three-body total angular momentum J$\leq \frac{15}{2}$. Such sets of partial waves
guarantee convergence of predictions for all observables, 
see~\cite{raport_stary,raport2005} for more technical details. 
The 3N bound states are obtained as in Ref.~\cite{Nogga}.

To estimate the theoretical errors of predictions arising from neglecting, at a given order
of the chiral expansion, the contributions from higher orders, we apply the prescription
given in~\cite{lenpic3, lenpic4}. Namely, we estimate the truncation error $\delta(X)^{(i)}$ of an observable $X$ at 
$i$-th order of the chiral expansion, with $i=0,2,3,\dots$. If $Q$ denotes the chiral expansion parameter,
the expressions for truncation errors are 
\begin{eqnarray}
\delta(X)^{(0)} &\geq& max \left( Q^2 \vert X^{(0)} \vert\,, \vert X^{(i \geq 0)} - X^{(j \geq 0)} \vert \right) \,,\nonumber \\
\delta(X)^{(2)} &=& max \left( Q^{3} \vert X^{(0)} \vert \,, Q \vert \Delta X^{(2)} \vert \,, \vert X^{(i \geq 2)} - X^{(j \geq 2)} \vert \right) \,, \nonumber \\
\delta(X)^{(i)} &=& max \left( Q^{i+1} \vert X^{(0)} \vert \,, Q^{i-1} \vert \Delta X^{(2)} \vert \,, Q^{i-2} \vert \Delta X^{(3)} \vert \right) \, {\rm for} \, i \geq 3\;.
\label{therrors}
\end{eqnarray}
In the above formulas $X^{(i)}$ is a prediction for the observable $X$ at $i$-th order, $\Delta X^{(2)} \equiv X^{(2)} - X^{(0)}$ and $\Delta X^{(i)} 
\equiv X^{(i)} - X^{(i-1)}$ for $i \geq 3$. We also require that $\delta(X)^{(2)} \geq Q \delta(X)^{(0)}$ and 
$\delta(X)^{(i)} \geq Q \delta(X)^{(i-1)}$ for $i \geq 3$.
In particular, such a way of theoretical error estimation
takes into account the fact that our present calculations are
incomplete as we do not include the 3N force starting from N$^2$LO.
Furthermore, we do \emph{not} estimate in this paper the uncertainty
from the truncation of the chiral expansion of the current operators,
which are included by means of the Siegert theorem at all considered
orders. The actual theoretical uncertainty may, therefore, be larger
than the values of $\delta (X)^{(i)}$ given below. A more reliable
uncertainty quantification requires performing complete calculations
including the corresponding 3N forces and exchange current operators.
This work is in progress.


\section{Results for deuteron photodisintegration}
\label{chap3}

\begin{figure}
\includegraphics[width=.8\textwidth,clip=true]{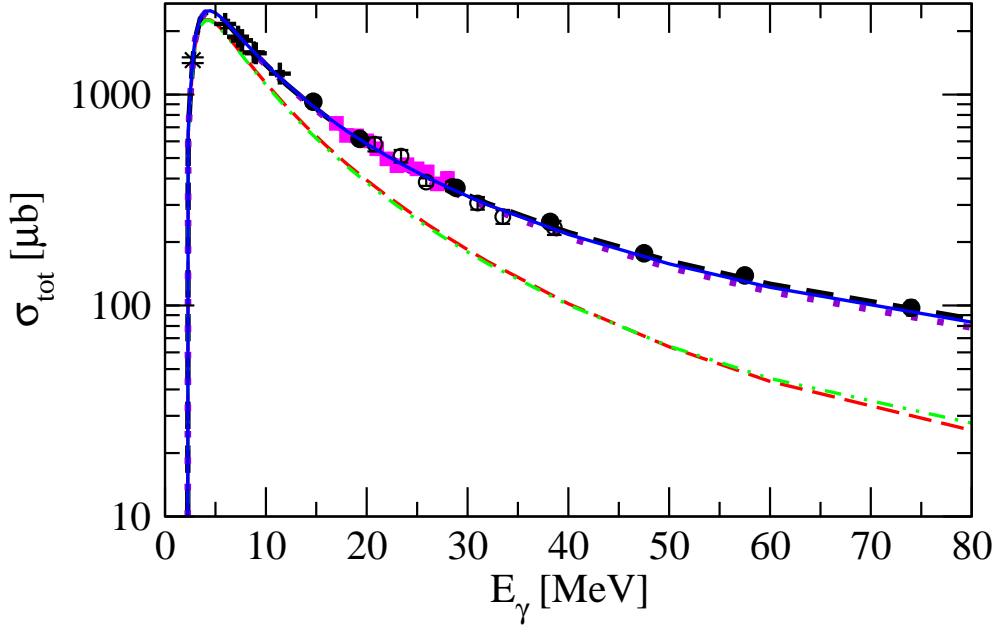}
\caption{(Color online) The total cross section $\sigma_{\rm tot}$ for the $\gamma + {\rm d} \rightarrow {\rm p} + {\rm n}$ reaction.
The chiral N$^4$LO, R=0.9 fm predictions for the SNC (SNC+Siegert) current model are shown with the dashed red (thick 
black dashed) curve.
The AV18 predictions for the SNC, SNC+Siegert and SNC+MEC current models are shown with the double-dotted-dashed green, dotted violet and
solid blue curve, respectively. The experimental data are from Ref.~\cite{Moreh} (black "x"),~\cite{Skopik} 
(magenta squares),~\cite{Bosman} (open circles), ~\cite{Birenbaum} (black pluses) and~\cite{Bernabei} (black dots).
}
\label{fig1}
\end{figure}

In Fig.~\ref{fig1} we show the total cross section for $\gamma + {\rm d} \rightarrow {\rm n} + {\rm p}$ process at 
photon laboratory energies E$_\gamma$ up to 80 MeV obtained using the chiral NN interaction at N$^4$LO~\cite{imp1, imp2} with the value of the 
regularization parameter $R$=0.9 fm. In this case we apply two models of the electromagnetic current:
the SNC alone (the dashed red curve) and the SNC+Siegert (the thick dashed black curve). 
It is clear that while the omission of 2N currents leads to a poor description of the data,
the SNC+Siegert approach yields an excellent agreement with the experimental results.
For the sake of comparison with the predictions
based on semi-phenomenological forces, we show also predictions obtained with the AV18 NN interaction~\cite{AV18}
and three models of the nuclear current: SNC (green double-dotted-dashed curve), SNC+Siegert (dotted violet curve) and
SNC+MEC (blue solid curve). The SNC+MEC model comprises single nucleon contributions 
and the explicit $\pi$-like and $\rho$-like meson exchange currents (MEC) linked to AV18 NN interaction
(see~\cite{Golak,raport2005} for more details). We observe that for this observable 
the implicit (SNC+Siegert) and explicit (SNC+MEC) ways of including many-body contributions 
to the current operator 
give quite similar predictions, which are in a very good agreement with the data,
when the  AV18 NN potential is used to generate the 2N states.
Further, the SNC+Siegert approach to the current operator works equally well 
with the chosen chiral and  the  AV18 NN potentials.
In both cases we obtain very similar predictions, practically indistinguishable at
photon energies below approximately 30~MeV. At the higher energies a small difference develops between the chiral and the AV18 potential, with the chiral predictions lying
closer to the data.

\begin{figure}
\includegraphics[width=1\textwidth,clip=true]{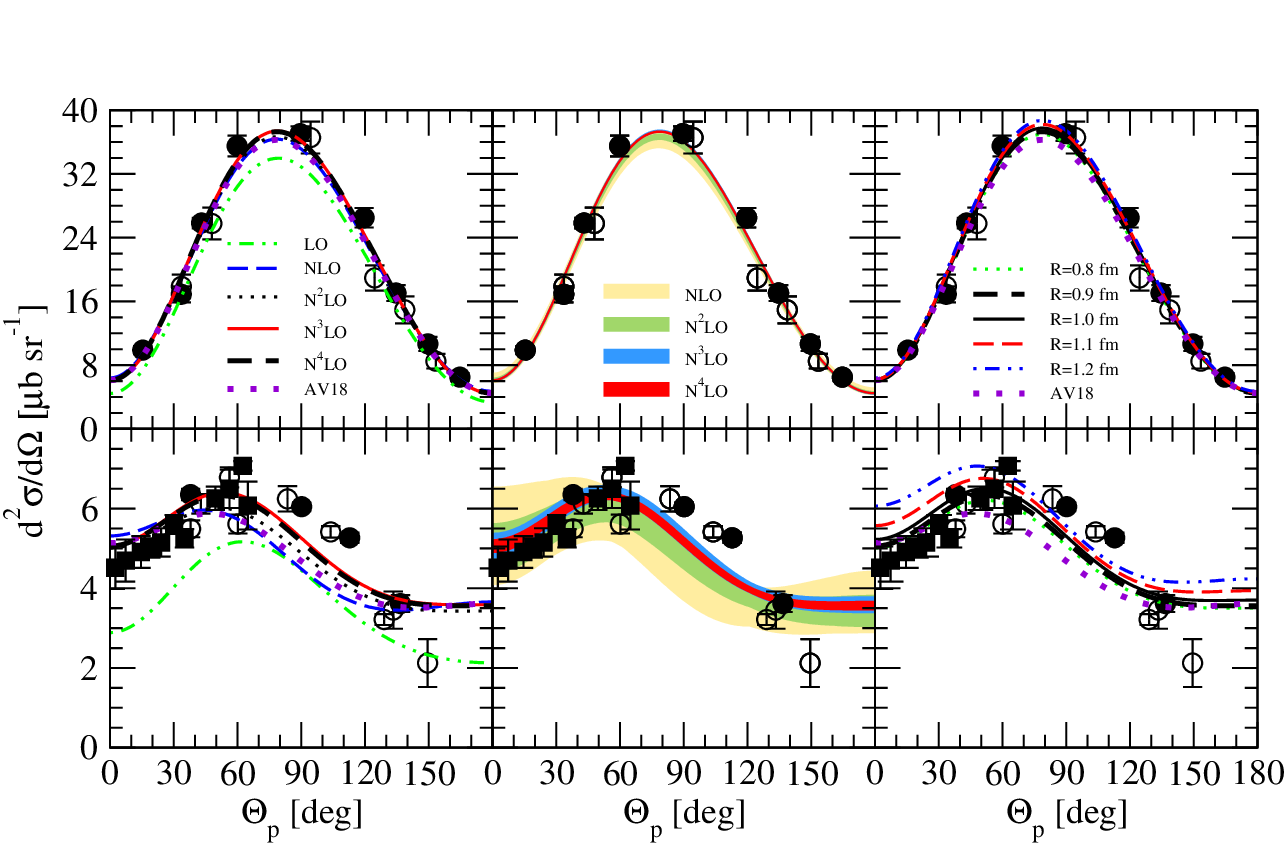}
\caption{(Color online) The differential cross section $\frac{{\rm d}^2\sigma}{{\rm d}\Omega}$ 
for the $\gamma + {\rm d} \rightarrow {\rm p} + {\rm n}$ reaction at 
E$_{\gamma}$=30~MeV (the upper row) and E$_{\gamma}$=100 MeV (the lower row) as a function 
of the proton c.m. scattering angle $\theta_{\rm p}$.
The left column shows the convergence of predictions at $R$=0.9 fm with respect to the order of the chiral expansion
(the double-dotted-dashed green, dashed blue, dotted black, solid red and thick dashed black curves 
correspond to predictions
at LO, NLO, N$^2$LO, N$^3$LO and N$^4$LO, respectively).
The middle column shows the truncation errors (see text) at the different orders of the chiral expansion:
NLO (yellow band), N$^2$LO (green band), N$^3$LO (turquoise band) and N$^4$LO (red band).
The right column shows the dependence of predictions at N$^4$LO on the value of parameter $R$ 
(the dotted green, thick dashed black, solid black, dashed red and double-dotted-dashed blue curves 
correspond to predictions 
with $R$=0.8, 0.9, 1.0, 1.1 and 1.2 fm, respectively). 
Note that the thick dashed black curve shown in both margin columns represents the same predictions. 
Also the thick violet dotted 
curve which represents the AV18 predictions is duplicated in the margin columns.
All data points (open and solid circles and full squares) are from~\cite{Ying}.
}
\label{fig2}
\end{figure}

\begin{figure}
\includegraphics[width=1\textwidth,clip=true]{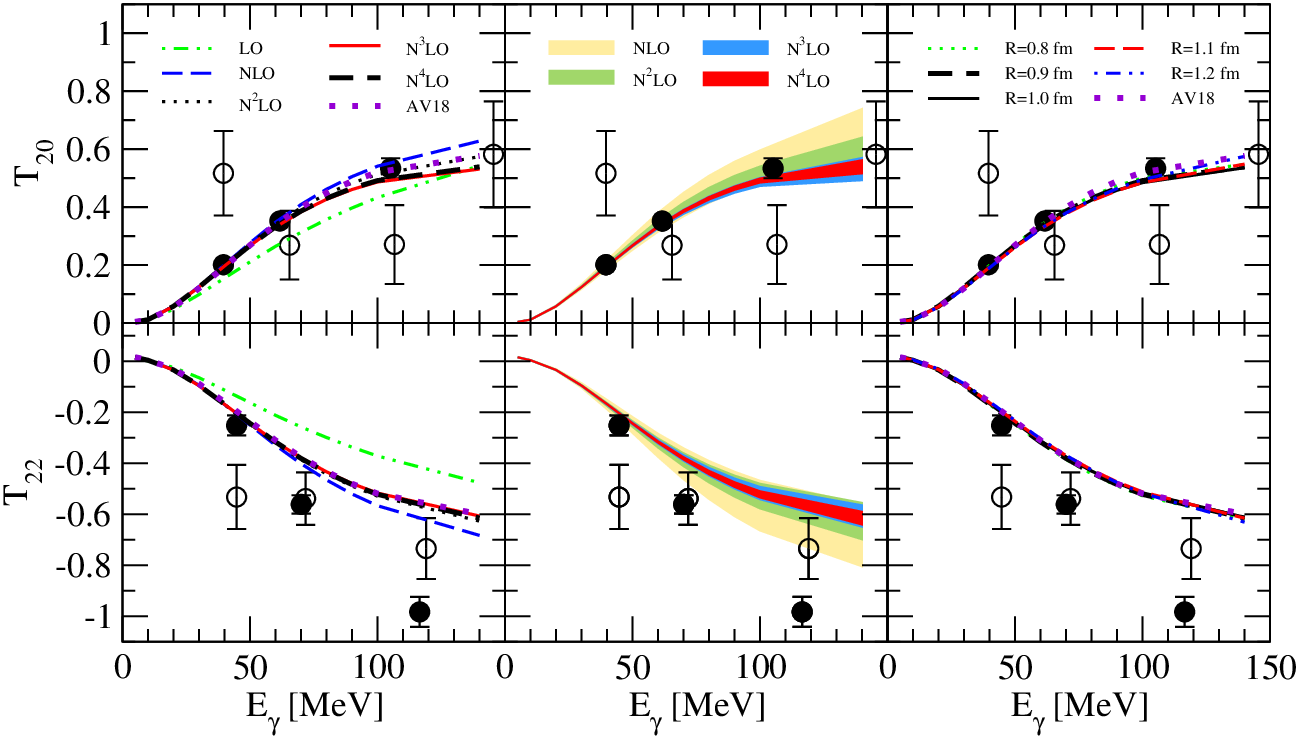}
\caption{(Color online) The deuteron analyzing powers T$_{20}$ (top) and T$_{22}$ (bottom) 
for the $\gamma + {\rm d} \rightarrow {\rm p} + {\rm n}$ reaction at $\Theta_{p}^{c.m.} = 
88^{\circ}$ and for E$_{\gamma}$ up to 150 MeV.
The left column shows the convergence of predictions at $R$=0.9 fm with respect to the order of the chiral expansion
(curves as in Fig.~\ref{fig2}).
The middle column shows the truncation errors (see text) at different orders of the chiral expansion
(bands as in Fig.~\ref{fig2}).
The right column shows the dependence of predictions at N$^4$LO on the value of the $R$ parameter
(curves as in Fig.~\ref{fig2}
). The data are from~\cite{Rachek} (filled circles) and~\cite{Mishnev} (open circles).
}
\label{fig3}
\end{figure}

\begin{figure}
\includegraphics[width=1\textwidth,clip=true]{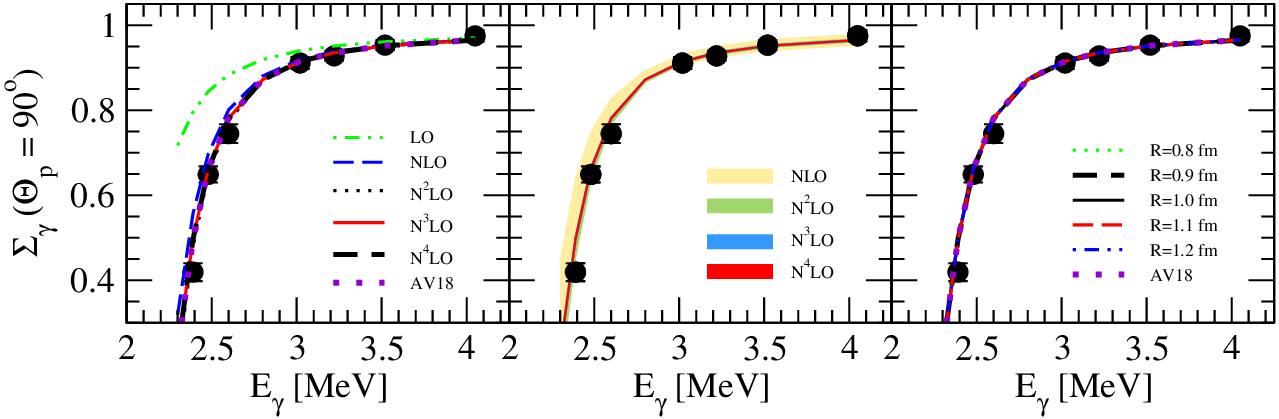}
\caption{(Color online) The photon asymmetry $\Sigma_{\gamma} (\theta_{\rm p}=90^o)$ at proton c.m. scattering angle
$\theta_{\rm p}=90^o$ 
for the $\gamma + {\rm d} \rightarrow {\rm p} + {\rm n}$ reaction at low photon energies.
The left column shows the convergence of predictions at $R$=0.9 fm with respect to the order of the chiral expansion
(curves as in Fig.~\ref{fig2}).
The middle column shows the truncation errors (see text) at the different orders of the chiral expansion
(bands as in Fig.~\ref{fig2}).
The right column shows the dependence of predictions at N$^4$LO on the value of the $R$ parameter
(curves as in Fig.~\ref{fig2}
). The data are from~\cite{Tornow}.
}
\label{fig4}
\end{figure}

\begin{figure}
\includegraphics[width=1\textwidth,clip=true]{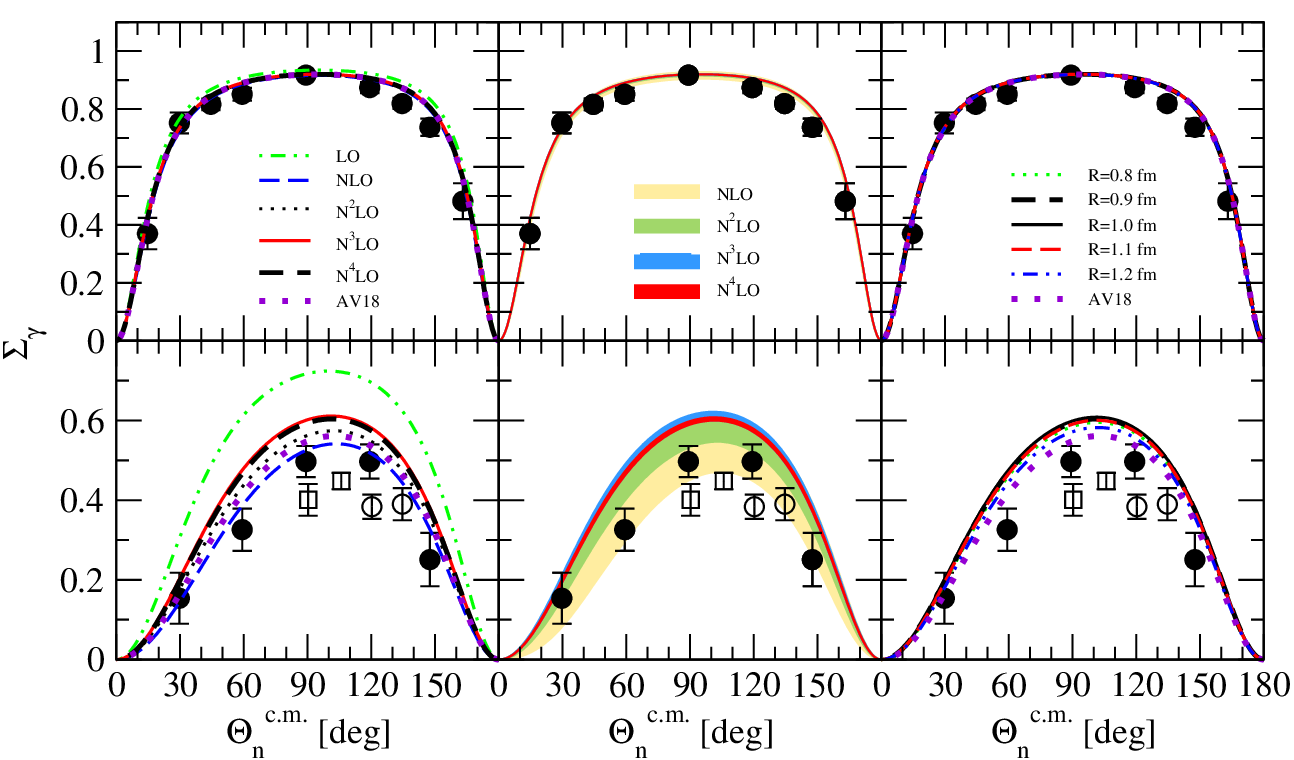}
\caption{(Color online) The photon asymmetry $\Sigma_{\gamma}$ for the $\gamma + {\rm d} \rightarrow {\rm p} + {\rm n}$ reaction 
at E$_{\gamma}$=19.8 MeV (top) and 60.8 MeV (bottom) as a function of neutron c.m. scattering angle $\theta^{c.m.}_{\rm n}$. 
The left column shows the convergence of predictions at $R$=0.9 fm with respect to the order of the chiral expansion
(curves as in Fig.~\ref{fig2}).
The middle column shows the truncation errors (see text) at the different orders of the chiral expansion
(bands as in Fig.~\ref{fig2}).
The right column shows the dependence of predictions at N$^4$LO on the value of the $R$ parameter
(curves as in Fig.~\ref{fig2}
). The data are from~\cite{Pascale} (filled circles),~\cite{Vnukov} (empty circles) and~\cite{Barannik} (squares).
}
\label{fig5}
\end{figure}

Next we study a more detailed observable, namely 
the differential cross section at two photon laboratory energies
E$_\gamma$=30~MeV (Fig.~\ref{fig2}, the upper row) and E$_\gamma$=100 MeV (Fig.~\ref{fig2}, the lower row).
In the left panel we show the convergence of predictions for $R$=0.9 fm with respect to the 
order of the chiral expansion. In the middle panel the uncertainty of
theoretical predictions due to the truncation of higher order contributions is given. Finally, in the right panel,
we demonstrate the dependence of predictions on the values of the regulator $R$ at N$^4$LO using
five different values of  $R$: 0.8, 0.9, 1.0, 1.1 and 1.2 fm.
Our best prediction, SNC+Siegert for $R$=0.9 fm is represented by the thick black dashed curve and is shown 
both in the left and right panels. For the sake of comparison, also the AV18 prediction given 
by the thick violet dotted line is displayed in these two panels.  
The same arrangement of curves will be preserved also in Figs.~\ref{fig3}-\ref{fig6},~\ref{fig7} and~\ref{fig11}.

It is clear that for both energies one has to go beyond the leading order (LO) to describe data. At the lower energy all the higher than LO
predictions are close to each other, but at E$_\gamma$=100 MeV the convergence is reached only at N$^3$LO. 
The truncation errors presented in the central panel confirm this observation and the band at N$^4$LO lies on the N$^3$LO one.
A small but visible width of the N$^4$LO band for the higher energy suggests that some contributions from higher orders are 
still possible for this observable. The cut-off dependence of the cross section is very small at lower energy and 
increases with energy, reaching at E$_\gamma$=100 MeV about 20\% at small proton c.m. scattering angles. However, a more careful analysis
reveals that predictions obtained with $R$=1.1 fm and $R$=1.2 fm are clearly far from the other ones, which are 
closer to each other. This observation is in agreement with the behaviour of the cross section for the NN and Nd elastic 
scattering~\cite{lenpic3,lenpic4} and provide yet another indication that the theoretical uncertainty of
the calculations using $R$=1.1~fm and $R$=1.2~fm is dominated by
finite-regulator artifacts, see Ref.~\cite{imp1} for more details.
Importantly, the cut-off dependence of the cross section observed here for the semi-local chiral force is much smaller
than the cut-off dependence observed for the older version of the potential with the nonlocal regularization.
As shown in Fig.~1 of Ref.~\cite{Rozpedzik} (pink band) for the older potential the cut-off dependence 
reaches 25\% already at E$_\gamma$=30~MeV and
increases with photon energy. 
For the improved chiral force at N$^4$LO all predictions are slightly above the cross section calculated with the AV18 potential.
The data at E$_\gamma$=100 MeV and at small angles between proton and photon momenta are
better described by the AV18 force while data at bigger angles are closer to the chiral predictions.

We choose the deuteron tensor analyzing powers T$_{20}$ and T$_{22}$ as examples of polarization 
observables. Such observables are supposed to be more sensitive to 
the NN interaction and the current operator used in the calculations. A measurement of these observables at
proton scattering angle $\theta_p^{c.m.}=88^{\circ}$ has been reported in Ref.~\cite{Rachek}.
In Fig.~\ref{fig3} we compare our predictions for T$_{20}$ (top) and T$_{22}$ (bottom) 
with precise data from Ref.~\cite{Rachek} and with older data from Ref.~\cite{Mishnev} 
for photon laboratory energies below E$_\gamma$=140 MeV.
The analyzing power T$_{20}$ is very well described by chiral predictions but for T$_{22}$
a clear discrepancy is seen with both sets of data above E$_\gamma$=50 MeV.
The predictions based on the chiral semi-local force show, for both analyzing powers, similar
behaviour as for the differential unpolarized cross section - their convergence with respect to the chiral expansion,
truncation errors and cut-off dependence are very reasonable. Even for the highest energies
the difference between N$^3$LO and N$^4$LO predictions is below 1.6\% (1.3\%) and the difference between predictions 
based on different values of  the $R$ parameter does not exceed 7.3\% (3.1\%) for T$_{20}$ (T$_{22}$). 
Note that if we omit predictions with $R$=1.2 fm, the latter numbers change to 2.3\% (1.4\%). 
The size of the truncation errors shows that only small contribution from higher orders can be expected 
even at energies above E$_\gamma$=50 MeV. 

It has been shown by Arenh\"ovel and collaborators
~\cite{Rachek, Arenhovel} in calculations with semi-phenomenological NN forces
that a more complete 2N current operator including 
in addition to the implicit MEC (incorporated in the Siegert approach) also
other explicit 2N operators, isobar configurations and leading order relativistic corrections
leads to a much better description of T$_{22}$. 
Thus, the poor description of T$_{22}$ data in Fig.~\ref{fig3} can be attributed to the weaknesses
of our model for the 2N electromagnetic current.
This is interesting in view of future studies which will be performed 
with NN interactions and current operators consistently derived within the $\chi$EFT framework.
We would like to stress that such studies would benefit from precise measurements of the deuteron 
analyzing powers at energies below E$_\gamma$=140 MeV in the whole range of scattering angles.

For the deuteron photodisintegration reaction also data for the photon asymmetry are available.
In a recent precision experiment~\cite{Tornow} the photon asymmetry $\Sigma_\gamma$ has been measured 
at the proton c.m. scattering angle $\theta=90^o$,  
for the low photon laboratory energies up to 4.05 MeV.
In Fig.~\ref{fig4} we compare our results with these data. As can be expected at such low energies even
predictions at lower orders are sufficient to describe the data. The results
at N$^2$LO, N$^3$LO and N$^4$LO practically overlap. Also the predictions for different
values of the $R$ regulator are very close to each other. The agreement with the data is excellent. 

A precise measurement of the photon asymmetry in the $\gamma + {\rm d} \rightarrow {\rm p} + {\rm n}$ reaction
as a function of the neutron c.m. scattering angle
at several photon energies has been reported in~\cite{Pascale}.
We choose two of them: E$_{\gamma}$=19.8 MeV and 60.8 MeV to give examples for 
small and medium photon energies. In Fig.~\ref{fig5} we compare our results with the data from~\cite{Pascale} and~\cite{Vnukov,Barannik}.
At the lower photon energy we observe a similar picture as for the photon asymmetry $\Sigma_{\gamma} (\theta_{\rm p}=90^o)$:
predictions are insensitive to the regulator value and to the order of the NN interaction. Even predictions at LO
describe data at E$_{\gamma}$=19.8 MeV reasonably well. At E$_{\gamma}$=60.8 MeV predictions for the photon asymmetry 
become convergent only at N$^3$LO.
The truncation errors and the range of predictions with the different values of the $R$
parameter at N$^4$LO remain small. This again can be compared with predictions for the nonlocal force~\cite{nucleon-nucleon}
shown in~\cite{Rozpedzik}. The improved interaction leads at E$_{\gamma}$=60.8 MeV to a theoretical uncertainty 
approximately two times smaller than for the older force. 
The chiral predictions at the lower energy are in agreement with the data and are slightly above them at the higher energy.
This may indicate that more sophisticated structures in the 2N current operator are required. 
The AV18 predictions are closer to the data
but still overpredict them at the maximum of the photon asymmetry.     

\section{Results for nucleon-deuteron radiative capture}
\label{chap4}

\begin{figure}
\includegraphics[width=1\textwidth,clip=true]{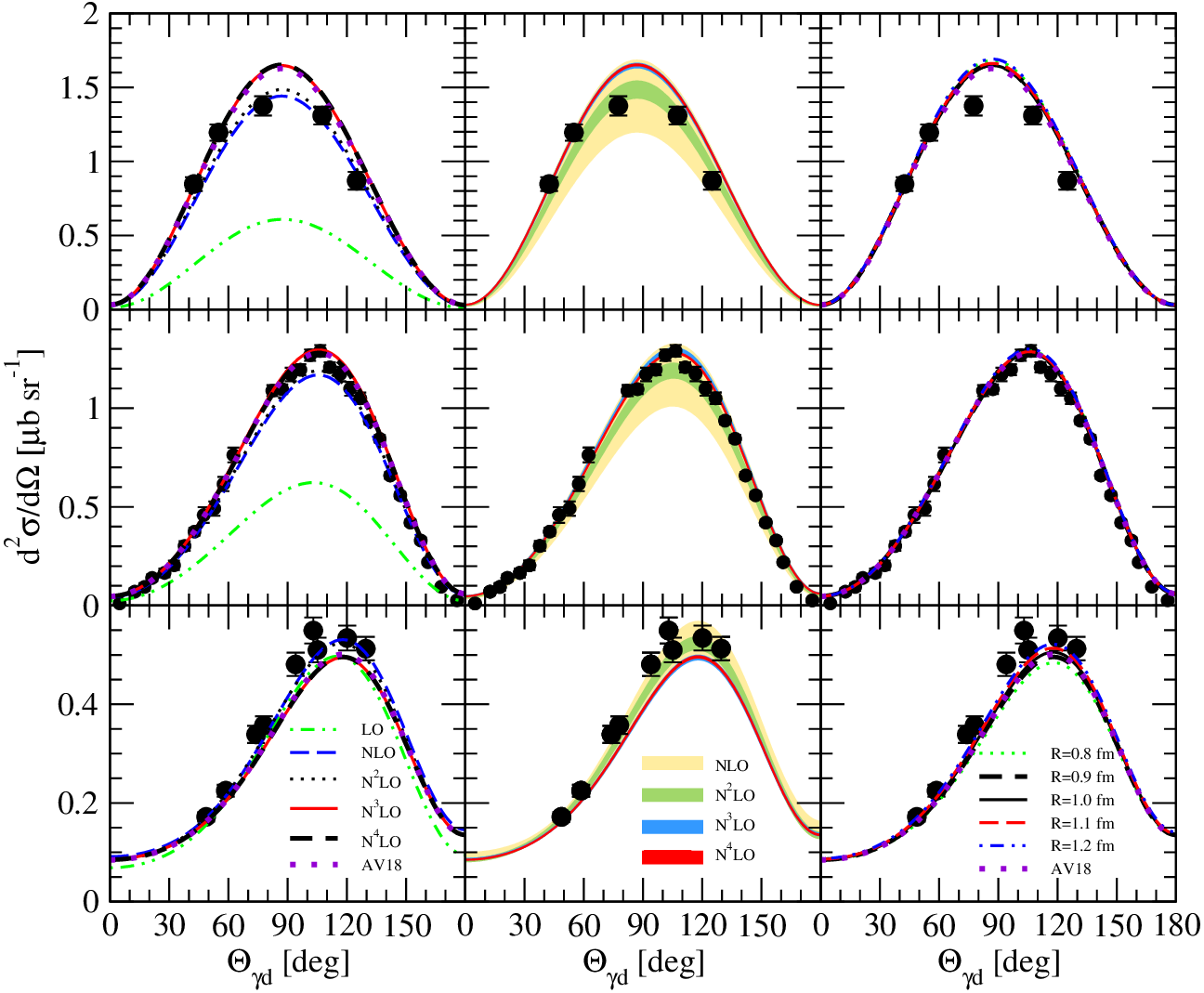}
\caption{(Color online) The differential cross section $d^2\sigma / d\Omega$ for the ${\rm n} + {\rm d} \rightarrow ^3{\rm H} + \gamma$ reaction
at E$_{\rm n}$=9.0 MeV (top) and for the ${\rm p} + {\rm d} \rightarrow ^3{\rm He} + \gamma$ reaction
at E$_{\rm p}$=29 MeV (middle) and E$_{\rm p}$=95 MeV (bottom).
The left column shows the convergence of predictions at $R$=0.9 fm with respect to the order of the chiral expansion
(curves as in Fig.~\ref{fig2}).
The middle column shows the truncation errors (see text) at the different orders of the chiral expansion
(bands as in Fig.~\ref{fig2}).
The right column shows the dependence of the predictions at N$^4$LO on the value of the $R$ parameter
(curves as in Fig.~\ref{fig2}
). The data at E$_{\rm n}$=9.0 MeV are from~\cite{Mitev}, at E$_{\rm p}$=29 MeV from~\cite{Belt} and 
at E$_{\rm p}$=95 MeV from~\cite{Pitts}.
}
\label{fig6}
\end{figure}

The three-nucleon systems pose another possibility to test models of nuclear dynamics.
In Fig.\ref{fig6} we show the differential cross section for the neutron-deuteron radiative capture reaction
at the neutron laboratory energy E$_{\rm n}$=9.0 MeV (upper row), and for the proton-deuteron radiative capture at
the proton laboratory energies E$_{\rm p}$=29.0 MeV (central row) and E$_{\rm p}$=95.0 MeV (lower row). 
For the neutron capture process the chiral predictions at next-to-leading order (NLO) and N$^2$LO agree with data but 
the N$^3$LO or N$^4$LO forces 
shift the predictions about 10\% above the experimental points. 
It was shown in~\cite{Skibinski_APP1} that for the case of nonlocal regularization 
the 3N force reduces the cross section for this process at low energies also by about 10\%. 
Thus it will be interesting to
check if the same is true for the locally regularized 3N force. 
The width of the band representing the truncation errors at N$^4$LO is small and 
the cross section is practically independent from the choice of the regulator value.

In the case of the proton-deuteron radiative capture at presented here proton energies the 
3N force effects (shown in Fig. 5 of Ref.~\cite{Skibinski_APP1}) are different.
At E$_{\rm p}$=29.0 MeV they are small, so it is very likely that agreement of N$^4$LO predictions with the data, observed
in Fig.\ref{fig6}, will remain, if the locally regularized 3N force is included.
At E$_{\rm p}$=95 MeV 3N force increases the cross section, so again it is possible that the observed underprediction 
of the cross section will be removed by the 3N interaction.
It seems rather accidental that even at the highest energy the LO predictions are relatively close to the other ones.
The truncation errors and cut-off dependence remain 
small for both proton energies. This is very different from results obtained with explicit chiral MEC at LO 
and dominant terms at NLO for the old version of
the potential with nonlocal regularization which show strong cut-off dependence - around
50\% at E$_{p}$=50 MeV (see Fig.4 of~\cite{Rozpedzik}). 
However, one has to be careful comparing predictions from Ref.~\cite{Rozpedzik} with the current ones since
it is expected that part of the cut-off dependence seen in Ref.~\cite{Rozpedzik} should be absorbed
into short-range currents, neglected in Ref.~\cite{Rozpedzik}.    
Nevertheless, much smaller cut-off dependence and truncation errors
observed with the local force~\cite{imp1,imp2} are very promising.

\begin{figure}
\includegraphics[width=1\textwidth,clip=true]{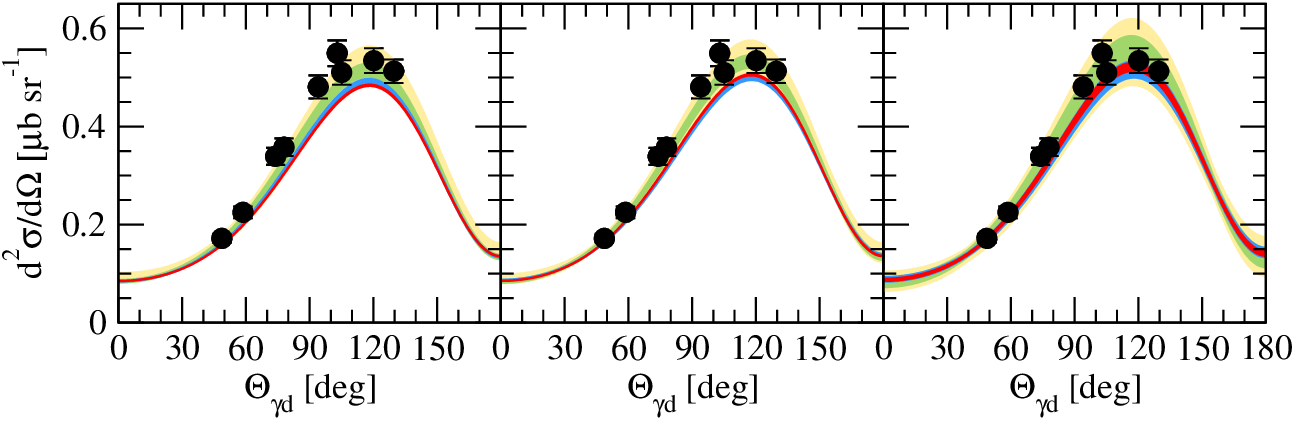} 
\caption{(Color online) The truncation errors at different orders of the chiral expansion for the same
cross section as shown in Fig.\ref{fig6} but at the proton energy E$_{\rm p}$=95 MeV only.
The predictions have been obtained using the value of the regularization parameter 
$R$=0.8 fm (left), $R=$1.0 fm (middle) and $R=$1.2 fm (right). 
Bands are as in Fig.~\ref{fig2} and data are from~\cite{Pitts}.  
}
\label{fig6a}
\end{figure}

The size of the truncation errors clearly depends on the value of the regularization parameter $R$.
In Fig.~\ref{fig6a} we compare the truncation errors for the same differential cross section as shown in the last row of 
Fig.\ref{fig6}, i.e. at E$_{\rm p}$=95 MeV but for $R$=0.8 fm (left), $R=$1.0 fm (middle) and $R=$1.2 fm (right). 
In all cases the truncation errors
decrease with the growing chiral order and are very small at N$^4$LO for $R$=0.8 and 1.0 fm, and much bigger for $R=$1.2 fm.
In particular, in the maximum of the cross section the bands width at N$^4$LO are 0.003 $\mu$b/sr both for $R$=0.8 
and 1.0 fm and 0.017 $\mu$b/sr for $R$=1.2 fm. The latter value is still 
approximately twice less than the spread of predictions at N$^4$LO obtained with different values of regularization 
parameter, which amounts $\Delta$=0.038 $\mu$b/sr in the maximum of the cross section.
This shows, that fixing the value of regulator parameter is important for the analysis of theoretical uncertainties.
In Refs.~\cite{imp1,imp2} the values of regularization parameter $R$=0.9 fm and $R$=1.0 fm have been 
recommended due to the best description of the nucleon-nucleon scattering data. Our findings on truncation errors 
also support this choice.

\begin{figure}
\includegraphics[width=1\textwidth,clip=true]{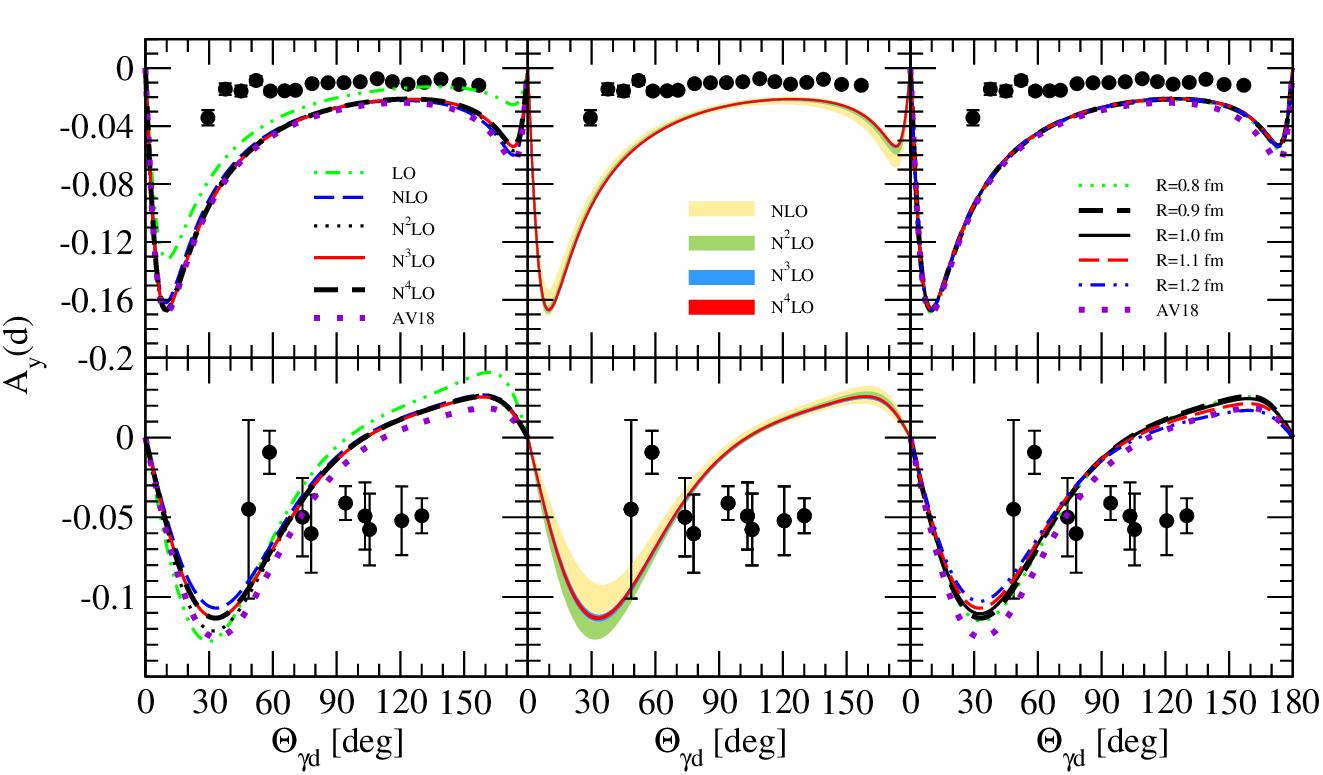}
\caption{(Color online) The deuteron analyzing power A$_y$(d) for the ${\rm p} + {\rm d} \rightarrow ^3{\rm He} + \gamma$ reaction
at the deuteron laboratory energies E$_{\rm d}$= 17.5 MeV (top) and E$_{\rm d}$=95 MeV (bottom).
The left column shows the convergence of the predictions at $R$=0.9 fm with respect to the order of the chiral expansion
(curves as in Fig.~\ref{fig2}).
The middle column shows the truncation errors at the different orders of the chiral expansion
(bands as in Fig.~\ref{fig2}).
The right column shows the dependence of predictions at N$^4$LO on the value of the $R$ parameter
(curves as in Fig.~\ref{fig2}
). The data at E$_{\rm p}$=17.5 MeV are from~\cite{Sagara, Akiyoshi} and at E$_{\rm p}$=95 MeV from~\cite{Pitts}.
}
\label{fig7}
\end{figure}

As an example of a polarization observable we choose the deuteron vector analyzing power A$_Y$
and show it in Fig.~\ref{fig7} at two deuteron laboratory energies E$_{\rm d}$= 17.5 MeV (top) and E$_{\rm d}$=95 MeV (bottom).
For both energies we observe nice behaviour at orders above N$^2$LO - the convergence with respect to chiral order 
is very good and
truncation errors are diminishing. Also the cut-off dependence is negligible. That is a significant 
improvement when compared to the case of
the nonlocal regularization~\cite{Skibinski_APP1}. At both energies the data description is poor, however 
this observable depends strongly on the details of the nuclear current operator~\cite{Golak}.  

\section{Three-body $^3$He photodisintegration}
\label{chap5}

\begin{figure}
\includegraphics[width=1\textwidth,clip=true]{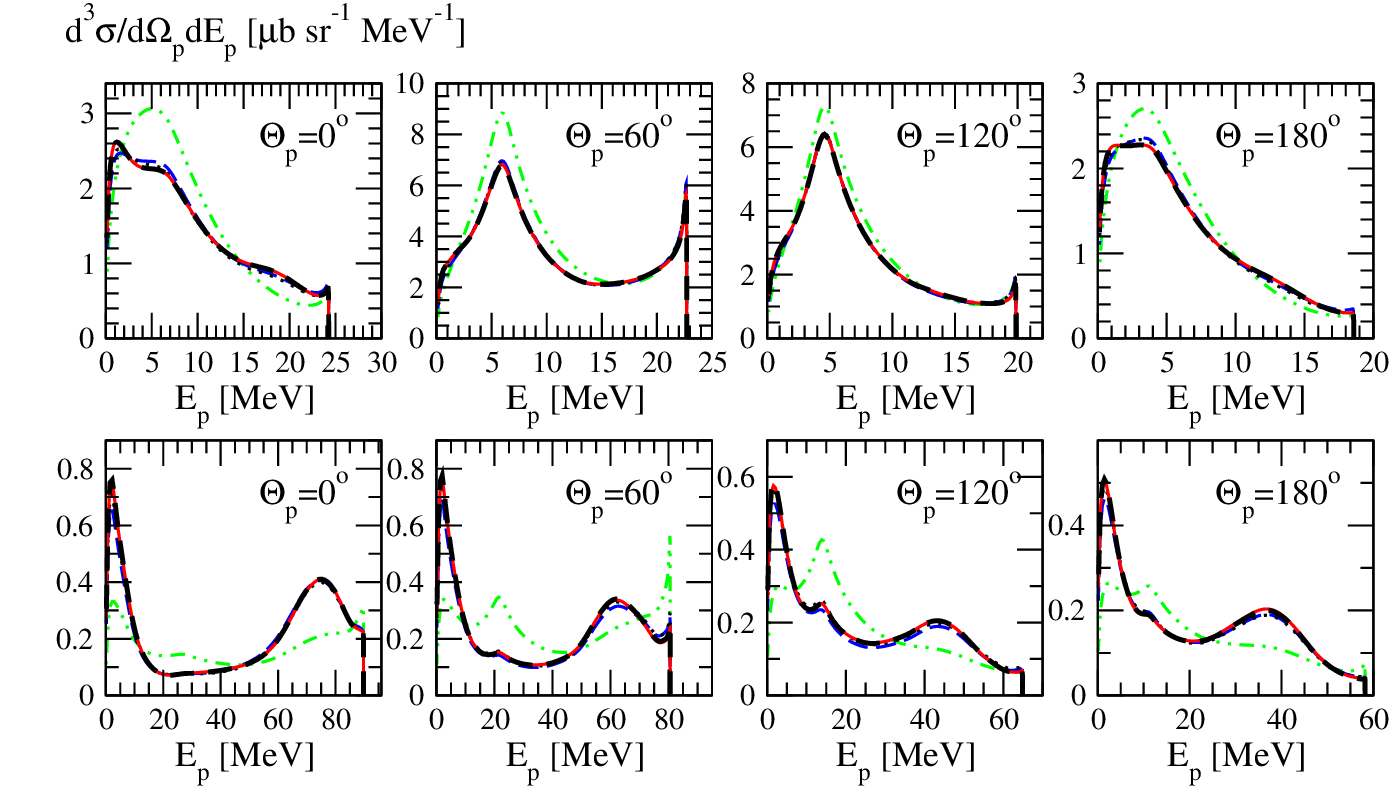}
\caption{(Color online) The semi-inclusive $^3{\rm He} (\gamma,{\rm p}){\rm p}{\rm n}pn$ cross section 
$\frac{d^3\sigma}{d\Omega_pdE_p}$ at E$_{\gamma}$=40 MeV (top) and E$_{\gamma}$=120 MeV (bottom), 
as a function of the outgoing proton energy E$_{\rm p}$ for various angles $\Theta_{\rm p}$ 
of the outgoing proton momentum with respect to the 
photon beam in the laboratory system. 
The predictions were obtained within the SNC+Siegert model and with the regulator $R=0.9$ fm. 
The double-dotted-dashed green, dashed blue, dotted black, solid red and thick 
dashed black curves correspond to LO, NLO, N$^2$LO, N$^3$LO and N$^4$LO predictions, respectively.}
\label{fig8}
\end{figure}

\begin{figure}
\includegraphics[width=1\textwidth,clip=true]{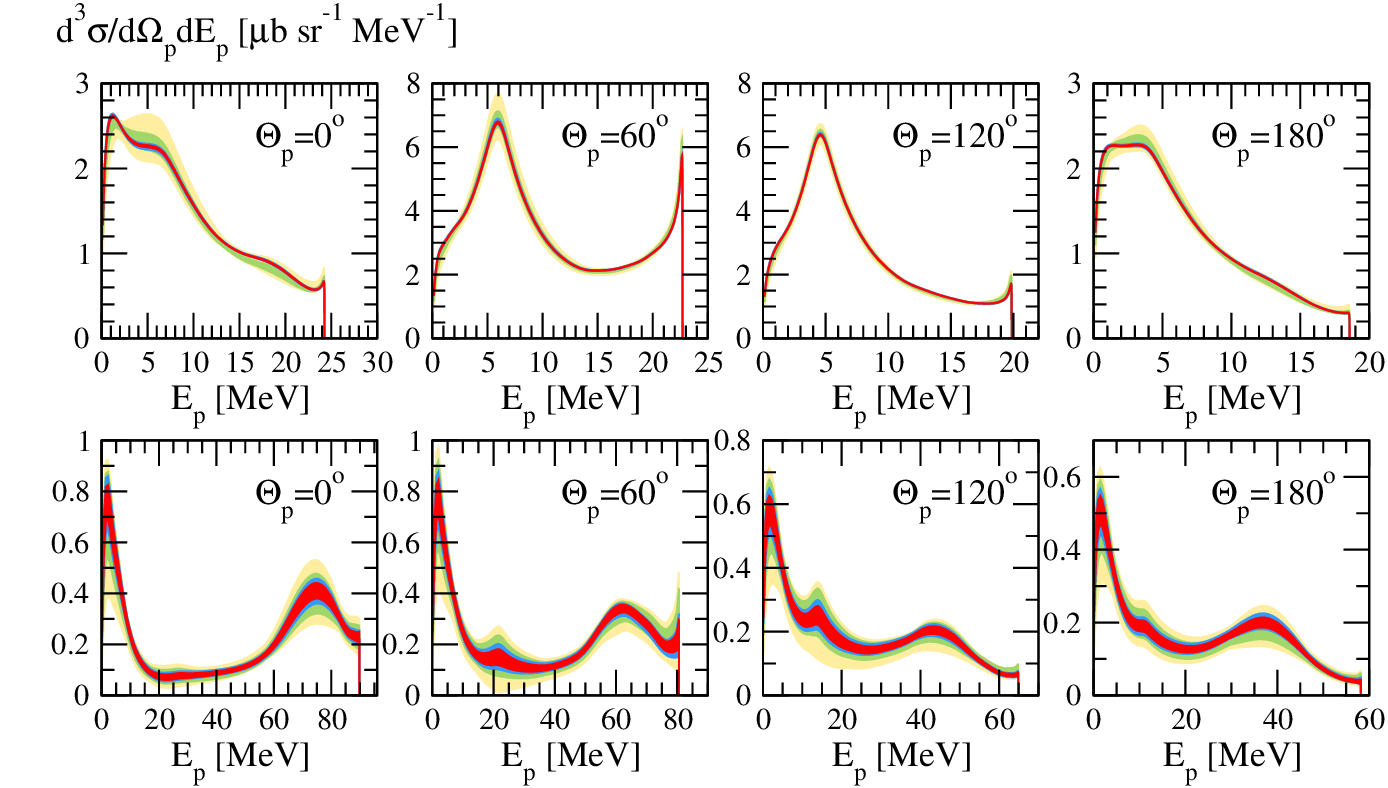}
\caption{(Color online) The estimated higher order truncation errors for the cross sections shown in Fig.~\ref{fig8}. 
The yellow, green, turquoise and red bands show the theoretical uncertainties 
at NLO, N$^2$LO, N$^3$LO and N$^4$LO, respectively.}
\label{fig9}
\end{figure}

\begin{figure}
\includegraphics[width=1\textwidth,clip=true]{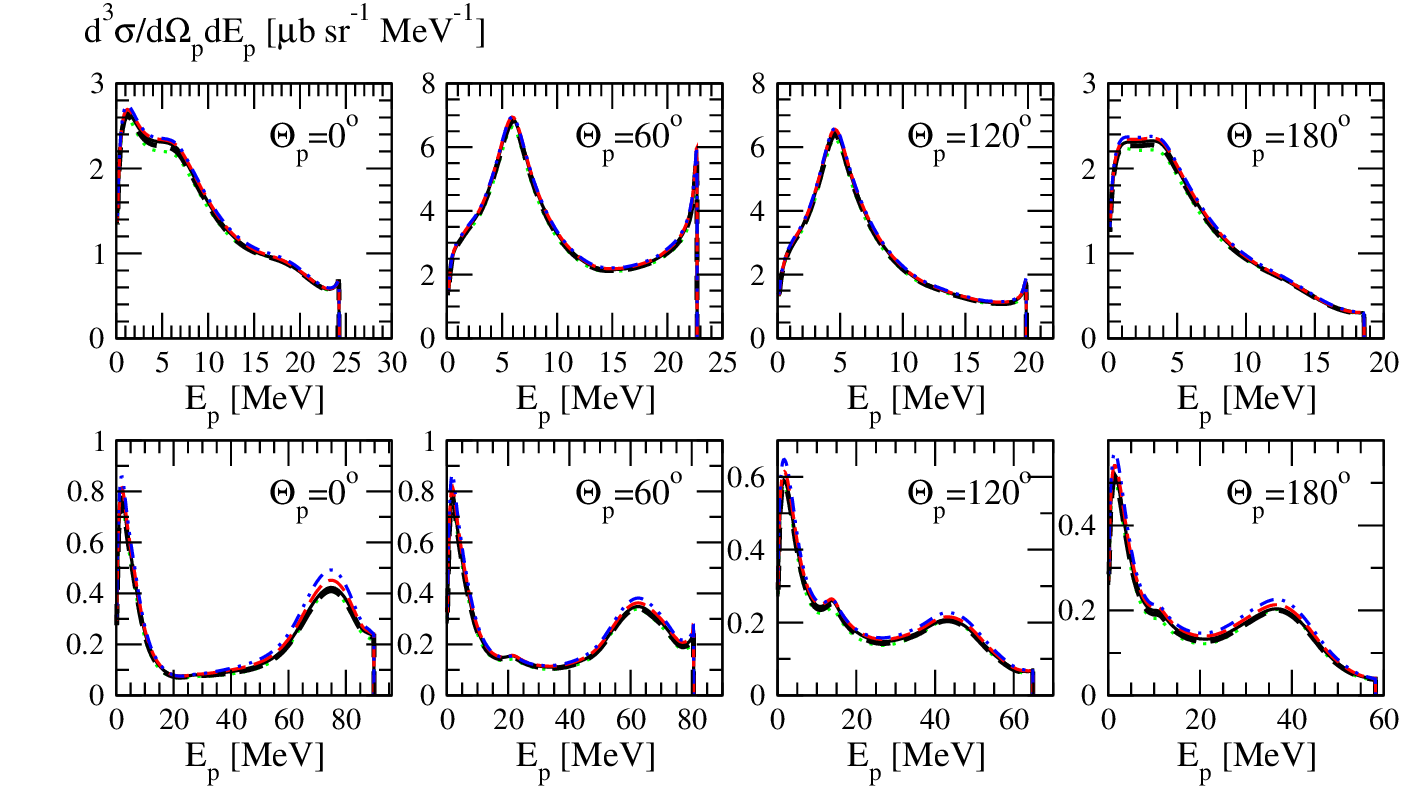}
\caption{(Color online) The cut-off dependence of the cross sections shown in Fig.\ref{fig8}.
The dotted green, thick dashed black, solid black, dashed red and double-dotted-dashed blue 
curves are for $R=0.8, 0.9, 1.0, 1.1, 1.2$ fm, respectively.
}
\label{fig10}
\end{figure}

We choose the semi-inclusive cross section $\frac{d ^ 3\sigma}{d\Omega_pdE_p}$ as an example of an observable
for three-body $^3$He photodisintegration.
In this process only one of the three outgoing nucleons is detected and we assume that it is a proton.
For this observables we have prepared three different Figs.~\ref{fig8}-\ref{fig10}, 
showing the dependence on the chiral order, error estimates and the dependence on the regulator value at N4LO.
In all three figures we show the cross section 
at photon laboratory energy E$_\gamma$= 40 MeV and 120 MeV 
as a function of the final proton energy for the proton emerging 
at four angles $\Theta_p$ with respect to the photon beam: 
$\Theta_{\rm p}= 0^{\circ}, 60^{\circ}, 120^{\circ}$ and $180^{\circ}$. 
Since we focus here on predictions of the new local chiral potential, we refer the reader to 
Refs.~\cite{gamma3N} and~\cite{raport2005} for the discussion on the origin of structures observed in the spectra.

In Fig.~\ref{fig8} we show the convergence of predictions with respect to the order of the chiral expansion
for the detected proton at E$_\gamma$= 40 MeV (top) and E$_\gamma$= 120 MeV (bottom).
Only predictions at LO are far away from the rest and are surely not sufficient to describe
the data. The other predictions are close to each other and, in particular
the N$^3$LO and N$^4$LO results, practically overlap.

The bands giving the truncation errors for the semi-inclusive cross section are
shown in Fig.~\ref{fig9}. At the photon laboratory energy E$_{\gamma}$=40 MeV a big 
contribution from higher orders is expected at the NLO (the yellow band) 
and still noticeable
addition is expected at 
N$^2$LO (the green band). At higher orders bands are very narrow and they practically overlap. 
Thus one can conclude that for the presented here cross section, calculations 
at N$^3$LO should be sufficient. At the higher photon laboratory energy E$_{\gamma}$=120 MeV the magnitude of the truncation 
errors is sizable even at N$^4$LO.   

Finally, in Fig.\ref{fig10} we explicitly show the dependence of the cross section on 
the value of the parameter $R$ used to regularize the chiral forces at N$^4$LO.
The cut-off dependence at E$_{\gamma}$=40 MeV is weak and its size is comparable with the
truncation errors. At the higher energy clear differences between predictions based on different values of $R$ are seen.
The range of predictions due to the different values of $R$ usually slightly exceeds the theoretical
uncertainties at N$^4$LO shown in Fig.~\ref{fig9}. Note, however, that 
the values $R$=0.9 fm and $R$=1.0 fm are preferred~\cite{imp1,imp2}.

\section{Muon capture}
\label{chap6}

For the muonic deuterium atom the weak capture process leads to 
two neutrons and a muon neutrino in the final state.
The total capture rates $\Gamma_{\rm d}$ for this process are given in Tab.~\ref{tab1}. We also give
the truncation error $\delta(\Gamma_{\rm d})^{(i)}$ (Eq.~\ref{therrors}) at a given order for $R$=0.9 fm (the next-to-last column) and
the maximal difference between predictions with different value of the regulator $R$ at a given order, 
$\Delta$ (the last column).
We applied the SNC+RC model of the current operator and used values of the regularization 
parameter $R$ in the range from  0.8 to 1.2 fm.
The truncation errors decreases significantly with the increasing order of the chiral expansion,
confirming nice convergence of capture rate. At N$^4$LO and $R$=0.9 fm $\delta(\Gamma_{\rm d})^{(5)}$ 
is about 0.02\%. The truncation errors for $R$=1.2 fm (not shown in the table) are much bigger, reaching 
0.18\% at N$^4$LO. 
The cut-off dependence is also weak, at N$^4$LO $\Delta$=1.7 s$^{-1}$ what 
amounts to about 0.44\% of the capture rate at $R$=0.9 fm. 
The new predictions are also in agreement with the result based on the AV18 NN force,
which for the same model of the weak current is 382.3 s$^{-1}$.

In the case of muon capture on $^3$He three different final states are possible.
First, the capture process can lead to the non-breakup channel with the final $^3$H nucleus and 
the outgoing neutrino. Results concerning the total 
capture rate for this process are given in Tab.~\ref{tab2}.
The LO predictions are far away from the
other results, but starting from NLO we see a nice convergence of results 
with respect to the order of the chiral expansion for all the values of $R$.
The truncation error $\delta(\Gamma)^{(0)}$ at LO is over two orders of magnitude bigger than the one
at N$^4$LO. Its value at N$^4$LO is approx. 0.07\% (0.52\%) of the capture rate using $R$=0.9 (1.2) fm.
Also in this case the dependence on the cut-off parameter is very weak.
The last column in Tab.~\ref{tab2} shows the maximal difference between predictions 
at a given order and for different regulators $R$, $\Delta$, which at N$^4$LO reaches approximately 1.76\% 
for $^3$He - $^3$H transition. The capture rate obtained for this process with the 
AV18 NN potential and using the 
same model of weak current (SNC+RC) equals 1295 s$^{-1}$ and is very close to the chiral value.

For both above-mentioned processes we observe that the cut-off dependence dominates the truncation errors if
all five considered here values of the regulator are taken into account in agreement with the arguments
given in~\cite{lenpic4}. If one restricts oneself 
only to the suggested values of the regulator ($R$=0.9 fm or $R$=1.0 fm) then the truncation errors
and the spread due to the different $R$ are of the same order. We also find, for both capture processes,
a small value of $\Delta$ at LO, which seems however to us somewhat accidental.
It is also interesting to notice that the total capture rates reach their maxima at 
the same values of the regulator $R$=0.9 fm and 1.0 fm. 
Finally, we emphasize that the truncation error $\delta (\Gamma_{\rm d})$
at N$^2$LO and higher orders, $\delta (\Gamma_{\rm d})^{(\geq 3)}$,
estimated using Eq. (2.6) does not include information about the actual
size of corrections $\Delta \Gamma_{\rm d}^{(\geq 3)}$, which is not
available from the incomplete calculations presented here. Also, the
uncertainty
from the truncation of the chiral expansion of the exchange current
operators is not taken into account. Thus, the obtained values of
$\delta (\Gamma_{\rm d})$ may underestimate the actual theoretical
uncertainty at higher orders. In the future, more complete calculations
will provide information on the size of contributions beyond N$^2$LO
which would allow us to perform a more reliable uncertainty quantification.

\begin{table}
\begin{tabular}{|c|c|c|c|c|c|c|c|}
\hline
chiral order & $R$=0.8 fm & $R$=0.9 fm & $R$=1.0 fm & $R$=1.1 fm & $R$=1.2 fm & $\delta(\Gamma_{\rm d})$ & $\Delta$ \\
\hline
LO &  396.0  & 397.4  & 398.4  & 398.9  & 399.2  & 21.02 & 3.2 \\
NLO &  384.2  & 385.8  & 387.2  & 388.6  & 389.8  & 4.84 & 5.6 \\
N$^2$LO &  385.0  & 386.1  & 387.2  & 388.3  & 389.3  & 1.11 & 4.3 \\
N$^3$LO  & 386.8  & 386.4  & 385.2  & 384.3  & 383.2  & 0.26 & 3.6 \\
N$^4$LO  & 385.5  & 386.1  & 386.3  & 385.6  & 384.6  & 0.06 & 1.7 \\
\hline
\end{tabular}
\caption{The doublet capture rates $\Gamma_{\rm d}$ in [s$^{-1}$] for the $\mu^- + {\rm d} \rightarrow {\rm n} + {\rm n} + \nu_{\mu}$
process obtained within the NN interaction at given order and the SNC+RC model of the nuclear weak current operator (see text).
In the next-to-last column the truncation error $\delta(\Gamma_{\rm d})$ at given order of the chiral expansion 
for the doublet capture rate, obtained for $R$=0.9 fm, is given.
In the last column the spread of the results at a given chiral order due to the different $R$ values,
$\Delta$ in [s$^{-1}$], is shown.
}
\label{tab1}
\end{table}

\begin{table}
\begin{tabular}{|c|c|c|c|c|c|c|c|}
\hline
chiral order & $R$=0.8 fm & $R$=0.9 fm & $R$=1.0 fm & $R$=1.1 fm & $R$=1.2 fm & $\delta(\Gamma)$ & $\Delta$ \\
\hline
LO &  1610 &  1618 &  1610 &  1594 &  1572 &  314.0 & 46 \\
NLO &  1330 &  1357 &  1381 &  1405 &  1427 & 72.2  & 97 \\
N$^2$LO &  1337 &  1356 &  1376 &  1395 &  1415 & 16.6 & 78 \\
N$^3$LO &  1314 &  1304 &  1289 &  1278 &  1266 & 3.8 & 48 \\
N$^4$LO &  1296 &  1307 &  1308 &  1299 &  1285 & 0.9 & 23 \\
\hline
\end{tabular}
\caption{The total capture rates $\Gamma$ in [s$^{-1}$] for the $\mu^- + ^3{\rm He} \rightarrow ^3{\rm H} + \nu_{\mu}$
process obtained within the NN interaction at given order and the SNC+RC model of the nuclear weak current operator (see text).
In the next-to-last column the truncation error $\delta(\Gamma)$ at i-th order of the chiral expansion
for the total capture rate, obtained for $R$=0.9 fm, is given.
In the last column the spread of the results at a given chiral order due to the different $R$ values,
$\Delta$ in [s$^{-1}$], is shown.
}
\label{tab2}
\end{table}

\begin{table}
\begin{tabular}{|c|c|c|c|c|c|c|c|}
\hline
chiral order & $R$=0.8 fm & $R$=0.9 fm & $R$=1.0 fm & $R$=1.1 fm & $R$=1.2 fm & $\delta(\Gamma)$ & $\Delta$ \\
\hline
LO      &  262 & 282 & 312 & 350 & 392 & 304.0 & 130 \\
NLO     &  536 & 525 & 515 & 504 & 492 & 69.9 & 44 \\
N$^2$LO &  547 & 539 & 529 & 518 & 507 & 16.1 & 40 \\
N$^3$LO &  584 & 586 & 592 & 596 & 603 & 3.7 & 19 \\
N$^4$LO &  590 & 584 & 583 & 587 & 595 & 0.9 & 12 \\
\hline
\end{tabular}
\caption{The same as in the Tab.~\ref{tab2} but for the $\mu^- + ^3{\rm He} \rightarrow {\rm d} + {\rm n} + \nu_{\mu}$
process.
}
\label{tab3}
\end{table}

\begin{table}
\begin{tabular}{|c|c|c|c|c|c|c|c|}
\hline
chiral order & $R$=0.8 fm & $R$=0.9 fm & $R$=1.0 fm & $R$=1.1 fm & $R$=1.2 fm & $\delta(\Gamma)$ & $\Delta$ \\
\hline
LO      &   95 &      99 &     105 &     113 &     120 & 70.0 & 26 \\
NLO     &  159 &     157 &     154 &     151 &     148 & 16.1 & 11 \\
N$^2$LO &  161 &     159 &     157 &     154 &     151 & 3.7 & 10 \\
N$^3$LO &  169 &     169 &     171 &     172 &     175 & 0.9 &  6 \\
N$^4$LO &  170 &     169 &     169 &     170 &     173 & 0.2 &  4 \\
\hline
\end{tabular}
\caption{The same as in the Tab.~\ref{tab2} but for the $\mu^- + ^3{\rm He} \rightarrow {\rm p} + {\rm n} + {\rm n} + \nu_{\mu}$
process.
}
\label{tab4}
\end{table}

\begin{figure}
\includegraphics[width=1\textwidth,clip=true]{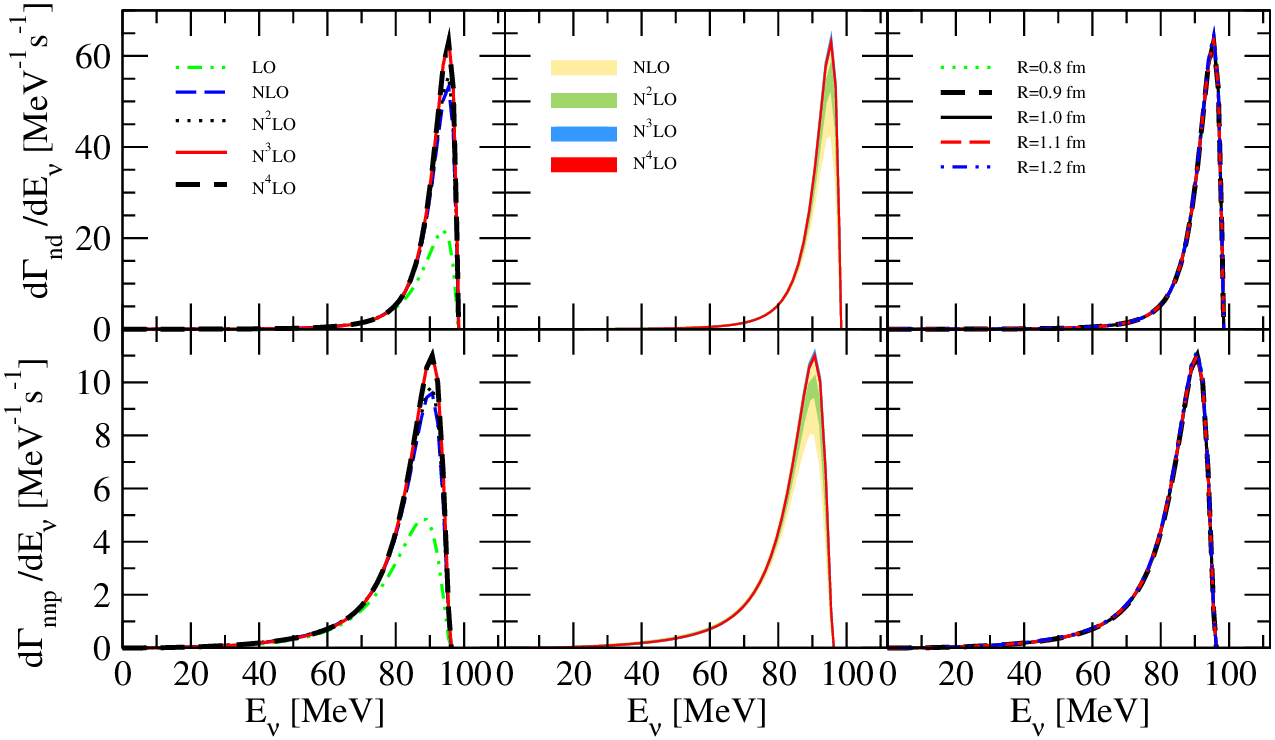}
\caption{(Color online) The differential capture rates for two- and three-nucleon breakup 
channels: $\frac{d\Gamma_{nd}}{dE_\nu}$ (top) and $\frac{d\Gamma_{nnp}}{dE_\nu}$ (bottom)
as a function of the outgoing neutrino energy E$_\nu$.
The left column shows the convergence of predictions at $R$=0.9 fm with respect to the chiral expansion order
(curves for chiral predictions only as in Fig.~\ref{fig2}).
The middle column shows the truncation errors at the different orders of the chiral expansion
(bands as in Fig.~\ref{fig2}).
The right column shows the dependence of predictions at N$^4$LO on the value of the $R$ parameter
(curves for chiral predictions only as in Fig.~\ref{fig2}
).
}
\label{fig11}
\end{figure}

In Fig.~\ref{fig11} we show the differential capture rates for the $\mu^- + ^3{\rm He} \rightarrow {\rm d} + {\rm n} + \nu_{\mu}$ (top) 
and $\mu^- + ^3{\rm He} \rightarrow {\rm p} + {\rm n} + {\rm n} + \nu_{\mu}$ (bottom) processes
as a function of the outgoing neutrino energy E$_{\nu}$.
For both processes the convergence of differential capture rates at $R$=0.9 fm with respect to the chiral order looks
similar. The NN force at LO underestimates the higher orders capture rates by a factor of approximately 3. 
The NLO and N$^2$LO predictions
are close to each other and finally the values at N$^3$LO and N$^4$LO are again very similar.
The contributions from orders beyond LO play a significant role only above E$_{\nu} \approx$80 MeV in both channels.
As in the case of observables for electromagnetic processes presented in the previous sections, the 
truncation errors are very small at the higher orders, pointing to a full convergence of nuclear forces with respect to
the chiral expansion. 
The cut-off dependence is small and predictions for different regulators overlap, 
except for the region
of the maximal capture rate values. However, even in this case the difference between predictions
based on the different $R$ values amounts only to 2.7\% (2.3\%) at E$_{\nu}$=95.5 (90.7) MeV for the two-body (three-body) channel.

The total capture rates for both breakup channels, given in Tabs.~\ref{tab3} and \ref{tab4} show 
similar behaviour as ones with the $^3$H in the final state, given in Tab.~\ref{tab2}. 
Both measures of theoretical uncertainties, $\delta(\Gamma)$ and $\Delta$, decreases with the chiral order,
and are below 2.5\% at N$^4$LO. Obtained here values of the capture rates are in agreement
with corresponding results~\cite{Golak_weak} for the AV18 force which are 604 s$^{-1}$ and 169 s$^{-1}$
for the n+d and n+n+p final states, respectively. 

\section{Summary}
\label{chap7}
We applied the recently developed improved chiral NN forces with the semi-local regularization~\cite{imp1,imp2} 
to a theoretical description of the deuteron and $^3$He photodisintegration reactions as well as 
to the proton-deuteron
radiative capture and the muon capture processes on the deuteron and $^3$He. 
The single nucleon electromagnetic current supplemented by implicit many-body contributions included via
the Siegert theorem was used for the processes with real photons. For the muon capture reactions
the single nucleon weak current operator was used supplemented with the dominant, p$^2$/M$_{\rm nucl}^2$ relativistic corrections.
Despite their simplicity such models of
nuclear currents are sufficient to realize the main goal of this work, which is to verify the usefulness 
of the locally regularized NN chiral forces in a description of electroweak processes at energies below the pion 
production threshold.

For all investigated reactions we could confirm the desired behavior of the used NN interaction.
Namely, we observed fast convergence of the predictions with respect to the order of
the chiral expansion - for all studied observables predictions at N$^4$LO are
very close to the ones at N$^3$LO. We also observe very weak dependence 
of our results on the value of the local regulator $R$. 
Especially, predictions with
the recommended values of regulator~\cite{imp1,imp2}, $R$=0.9 fm and $R$=1.0 fm, usually overlap at N$^4$LO.
The observed cut-off dependence is much weaker than the one found in Ref.~\cite{Skibinski_APP1} for 
the older chiral forces~\cite{nucleon-nucleon}, which were regularized directly in momentum space 
using nonlocal regulators. Finally, we estimate the truncation errors 
coming from neglecting higher order contributions.
These theoretical uncertainties decrease with the chiral order and are very small beyond N$^2$LO,
except for the highest energies studied here. The magnitude of the truncation errors depends on the
value of the regulator $R$ and is the biggest for $R$=1.2 fm. The sizes of the truncation errors at $R$=0.9 fm and $R$=1.0 fm
are comparable; they are of the same order as the difference between predictions obtained using these regulators.
When all the $R$ values are taken into account, the spread of predictions is usually bigger than
the truncation error, even that for $R$=1.2 fm. We conclude that in the future only recommended values
$R$=0.9 fm and $R$=1.0 fm should be used. However, it would be interesting to confirm this observation in 
nuclear structure calculations.

The quality of the description of the data is rather satisfactory but leaves room for improvement by
including contributions 
from 3N forces and many-body parts of the nuclear current operators. Thus, it will be important
to identify observables which are sensitive to the details of the dynamical framework. One of the candidate
is the deuteron analyzing power T$_{22}$ in the deuteron photodisintegration process.
Experimental efforts focused on precise and systematic measurements of such observables would 
be very important to provide a solid basis for a detailed study of chiral dynamics.

We may thus conclude that the present work confirms the importance of the improved chiral NN
potential with the local regularization for few-body investigations
in a broad range of energies.
Of course we are aware that the final conclusions about the observed patterns
can be drawn only when consistent 3N force and current operators at all the considered orders of the chiral expansion are included,
but the predictions presented here constitute a promising, inescapable first step in this direction.

\acknowledgments
Authors would like to thanks the members of the LENPIC collaboration for discussions.
This work was supported by the Polish National Science Center under Grants No. DEC-2013/10/M/ST2/00420.
and DEC-2013/11/N/ST2/03733, the ERC Starting Grant 259216 NuclearEFT, the DFG grant
SFB/TR 16 "Subnuclear Structure of Matter" and by the Chinese Academy of Science (CAS) President's
International Fellowship Initiative (PIFI) (Grant No. 2015VMA076). 
The numerical calculations have been performed on the supercomputer
cluster of the JSC, J\"ulich, Germany.

\end{document}